# Incorporating Affect in an Engineering Student's Epistemological Dynamics

Brian A. Danielak, Ayush Gupta, Andrew Elby
Departments of Curriculum and Instruction and Physics, University of Maryland
briandk@umd.edu, ayush@umd.edu, elby@umd.edu

**Abstract:** Research has linked a student's affect to her epistemology (Boaler & Greeno, 2000), but those constructs often apply broadly to a discipline and/or classroom culture. Independently, an emerging line of research shows that a student in a given classroom and discipline can shift between multiple locally coherent epistemological stances (Hammer, Elby, Scherr, & Redish, 2005). Our case study of Judy, an undergraduate engineering major, begins our long-term effort at uniting these two bodies of literature.

## Introduction

In this poster, we incorporate an affective node into a fine-grained toy model of an engineering major's cognitive dynamics during a clinical interview. We seek to characterize Judy's annoyance toward a particular kind of homework problem in her Circuits course, and its consequences. In our model, affect is not just an add-on; we use it to account for aspects of her conceptual and epistemological thinking that epistemology-only models can't and don't explain.

This kind of modeling responds to a specific literature gap between emotion and cognition. Many well-known studies link learners' affect and epistemology with large, often group-level constructs (Boaler & Greeno, 2000; Duit & Treagust, 2003; Lee, Hansen, & Wilson, 2006; Pintrich, Marx, & Boyle, 1993). Those constructs provide broad descriptive power, but tell us less about contextualized, in-the-moment student behaviors. Conversely, sensitive accounts of students' conceptual and epistemological cognition are plentiful in fine-grained details (diSessa, 1993, 2006; Hammer et al., 2005), but sparing in relating emotion to conceptions and epistemologies.

## Summary of Three Patterns of Judy's Thought and Behavior

In a one-hour clinical interview, we probed Judy's views about her Circuits course and her approaches toward problem-solving. Working as a group we then identified four cognitive patterns in her responses to certain prompts, seeking confirmatory or non-confirmatory evidence elsewhere in the data (Miles & Huberman, 1984).

*(1) Annoyance at conceptual problems:* Like many students in her class, Judy distinguished between two types of homework and exam problems: traditional equation-based problem-solving vs. conceptual problems asking for explanations of what's going on physically. Judy repeatedly emphasized that conceptual problems were *annoying* in words ("they are kind of annoying"), gestures, and facial expressions (frowns and facial distortions).

*(2) Conceptual reasoning as useless:* Judy sees conceptual reasoning as useless for practical purposes and considers equation-based problem-solving as "more helpful" and "more important."

*(3) Real/ideal gulf:* In Judy's view, her course spends too much time discussing "ideal circuits and theoretical methods," which "are not related to the actual circuit" or to "a professional engineer's job." This unbridgeable gulf between ideal and real circuits, she says, is what makes the conceptual problems annoying; those problems call for students to use concepts that apply only to ideal circuits.

*(4) Non-activation of (3) during traditional problem-solving.* When talking about or engaging in equation-based problem solving, Judy displays positive affect, and we find no evidence that in these moments she is ever thinking about the real/ideal gulf. Crucially, the equations she uses encode the same idealizing assumptions she disparages elsewhere in the interview. Her epistemological view about the real/ideal gulf is thus context-dependent: present when she *discusses* conceptual problems, but absent when she *solves* traditional ones.

## The Need for Affect in Toy Models of Judy's in-the-moment cognition

The four patterns discussed above are consistent with a context-dependent but affect-free toy model (Fig 1 (a)) in which Judy's epistemological view about the gulf between real and ideal circuits reinforces and is reinforced by her sense that the conceptual problems are useless. But two other episodes from the interview show the inadequacy of this model.

*Disconfirmatory episode A.* Near the end of the interview, Judy builds momentum using conceptual reasoning to (re)solve a homework problem from earlier in the course. She thinks in terms of "squeezing" the area under a voltage vs. time curve to change the peak voltage without affecting the average voltage. In her successful solving, she shows positive affect. Afterward, the interviewer asks about the squeezing: "Do you ever

…think about the mathematics that you use in that way?" Judy laughs, says "no no no," and adds "No, I never think of that before…. Yeah, I mean it's I feel like it's not very formal [smiles], but it's very useful." The toy model in figure 1 (a) doesn't predict such a reversal of her epistemological stance towards conceptual reasoning.

*Disconfirmatory episode B.* At one point, not discussed above, while Judy was still thinking about her annoyance at the conceptual problems, the interviewer probes her views about *both* the real/ideal gulf *and* conceptual reasoning: "So do you think if you're analyzing a real-world circuit, it's important to know about the physical aspects of the circuit?" Judy responds, "Not very important." The two-node model in figure 1 (a) didn't predict this.

To explain these two episodes *and* the four patterns discussed above, we must modify our toy model to include an affective state, Judy's *annoyance* at conceptual problems, as mediating the connection between her epistemological stance and her view that conceptual problems are useless (Fig. 1(b)). Our proposed three-node toy model explains how suppressing *annoyance* can cue Judy's epistemological reversal in Episode A. It also explains explains Episode B, because targeting Judy's epistemology (by focusing on *real* circuits) without addressing Judy's affect fails to suppress Judy's view that conceptual reasoning is "not very important." Ultimately, though limited to one student in one interview, our analysis shows the importance and feasibility of incorporating affect into fine-grained models of cognitive dynamics.

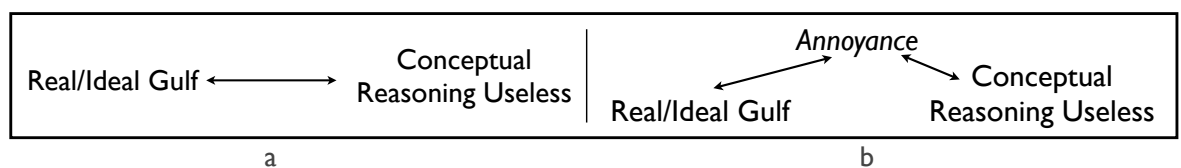

Figure 1 - Toy cognitive model with (a) only epistemological nodes and (b) incorporating affect (annoyance) with epistemology. The model in (b) better explains the contextual dynamics of Judy's epistemological stances.

## Acknowledgments

The authors wish to thank Eric Kuo and Michael M. Hull for their insights and analysis. Without them this project would not have been possible. The authors also wish to think Jennifer Richards and Laura Cathcart for feats of prestidigitation and writing support. This work was partially supported by NSF-EEC grant # 0835880 and NSF-DRL grant # 0733613.

## References

Boaler, J., & Greeno, J. G. (2000). Identity, agency, and knowing in mathematical worlds. In J. Boaler (Ed.), *Multiple Perspectives on Mathematics Teaching and Learning*, International perspectives on mathematics education (pp. 171-200). Westport, CT: Ablex Pub.

diSessa, A. A. (1993). Toward an Epistemology of Physics. *Cognition and Instruction*, *10*(2/3), 105-225.

diSessa, A. A. (2006). A history of conceptual change research: Threads and fault lines. In R. K. Sawyer (Ed.), *The Cambridge Handbook of the Learning Sciences* (pp. 265-282). Cambridge: Cambridge University Press.

Duit, R., & Treagust, D. F. (2003). Conceptual change: a powerful framework for improving science teaching and learning. *International Journal of Science Education*, *25*(6), 671-688. doi:Article

Hammer, D., Elby, A., Scherr, R. E., & Redish, E. F. (2005). Resources, Framing, and Transfer. In J. P. Mestre (Ed.), *Transfer of Learning from a Modern Multidisciplinary Perspective*, Current perspectives on cognition, learning, and instruction. Greenwich, CT: IAP.

Lee, L. A., Hansen, L. E., & Wilson, D. M. (2006). The impact of affective and relational factors on classroom experience and career outlook among first-year engineering undergraduates. Presented at the 36th ASEE/IEEE Frontiers in Education Conference. Retrieved from http://www.fie-conference.org/fie2006/papers/1796.pdf

Miles, M. B., & Huberman, A. M. (1984). *Qualitative Data Analysis: A Sourcebook of New Methods*. Beverly Hills: Sage Publications. Retrieved from http://lccn.loc.gov/84002140

Pintrich, P. R., Marx, R. W., & Boyle, R. A. (1993). Beyond Cold Conceptual Change: The Role of Motivational Beliefs and Classroom Contextual Factors in the Process of Conceptual Change. *Review of Educational Research*, *63*(2), 167-199.